\documentclass{article}

\usepackage{arxiv}
\usepackage[utf8]{inputenc} 
\usepackage[T1]{fontenc}    
\usepackage{hyperref}       
\usepackage{url}            
\usepackage{booktabs}       
\usepackage{amsfonts}       
\usepackage{amsmath}        
\usepackage{bm}
\usepackage{nicefrac}       
\usepackage{microtype}      
\usepackage{cleveref}       
\usepackage{lipsum}         
\usepackage{graphicx}
\usepackage[square,sort,comma,numbers]{natbib}
\usepackage{doi}
\usepackage[labelfont=bf]{caption}
\usepackage{float,lscape}
\usepackage[nottoc,numbib]{tocbibind}
\usepackage{caption}        
\usepackage[ruled,vlined]{algorithm2e}
\usepackage{subcaption}
\usepackage{scalerel}
\usepackage{xcolor}
\usepackage[width=.9\textwidth]{caption}

\title{Variational Bayes latent class approach for EHR-based phenotyping with large real-world data}

\date{}

\author{
Brian Buckley
\thanks{School of Mathematics \& Statistics, University College Dublin, Ireland}\\
\texttt{brian.buckley.1@ucdconnect.ie}
\And
Adrian O'Hagan
\footnotemark[1] \thanks{The Insight Centre for Data Analytics, University College Dublin, Ireland}
\And
Marie Galligan
\thanks{School of Medicine, University College Dublin, Ireland}
}

\renewcommand{\headeright}{}
\renewcommand{\undertitle}{}

\hypersetup{
pdftitle={Variational Bayes latent class approach for EHR-based phenotyping},
pdfsubject={q-bio.NC, q-bio.QM},
pdfauthor={Brian Buckley, Adrian O'Hagan, Marie Galligan},
pdfkeywords={Bayesian Latent Class Analysis, Variational Inference},
}

\begin{document}
\maketitle

\begin{abstract}
Bayesian approaches to clinical analyses for the purposes of patient phenotyping have been limited by the computational challenges associated with applying the Markov-Chain Monte-Carlo (MCMC) approach to large real-world data. Approximate Bayesian inference via optimization of the variational evidence lower bound, often called Variational Bayes (VB), has been successfully demonstrated for other applications. We investigate the performance and characteristics of currently available R and Python VB software for variational Bayesian Latent Class Analysis (LCA) of realistically large real-world observational data.  We used a real-world data set, Optum\textsuperscript{TM} electronic health records (EHR), containing pediatric patients with risk indicators for type 2 diabetes mellitus that is a rare form in pediatric patients. The aim of this work is to validate a Bayesian patient phenotyping model for generality and extensibility and crucially that it can be applied to a realistically large real-world clinical data set.  We find currently available automatic VB methods are very sensitive to initial starting conditions, model definition, algorithm hyperparameters and choice of gradient optimiser. The Bayesian LCA model was challenging to implement using VB but we achieved reasonable results with very good computational performance compared to MCMC.
\end{abstract}


\section{Introduction}

With growing acceptance by clinical regulators of using real-world evidence to supplement clinical trials, there is increasing interest in the use of Bayesian analysis for both experimental and observational clinical studies \cite{boulanger703and}.  Secondary use of EHR data has long been incorporated in phenotypical studies \cite{hripcsak2013next}.  The most popular approach that has evolved is a rules-based heuristic driven by data availability for a given phenotypic study.  Bayesian statistics provides a formal mathematical method for combining prior information with current information at the design stage, during the conduct of a study, and at the analysis stage.  This interest has been limited by the computational challenges of applying the Markov-Chain Monte-Carlo (MCMC) approach to large real-world clinical data.  MCMC is considered the gold standard for Bayesian inference because it asymptotically converges to the true posterior distribution.  The variational approach minimises the distance between a postulated family of standard distributions to approximate the posterior distribution, rather than directly sampling from it as done under MCMC.  This optimisation approach is often significantly more computationally efficient than MCMC at the expense of an estimated posterior.  We therefore need to carefully assess the quality of the variational approach if it is to be a credible method in the clinical context.

\vspace{5mm}
Latent Class Analysis is widely used where we want to identify patient phenotypes or subgroups given multivariate data \cite{lanza2013latent}.  A challenge in clinical LCA is the prevalence of mixed data, where we may have combinations of continuous, nominal, ordinal and count data together with missing data across variables.  Bayesian approaches to LCA may better account for this data complexity.  The Bayesian approach bridges rule-based phenotypes, which rely heavily on expert knowledge and opinion, and data-driven machine learning approaches, which are derived entirely from information contained in the data under investigation with no opportunity to apply relevant prior information.  We consider the common clinical context where gold-standard phenotype information is not available, such as genetic and laboratory data.  A general model for this context has high potential applicability across disease areas and both primary and secondary clinical databases for use in clinical decision support.  We used the same Bayesian LCA model form and motivating example as Hubbard et al. \cite{hubbard2019bayesian}, namely pediatric type 2 diabetes mellitus (T2DM), in order to test if the proposed LCA model translates to another electronic health records (EHR) database with the same target disease under study.  Paediatric T2DM is rare so it naturally gives rise to the data quality issues we would like to explore with our approach.

\section{Data}

This study used Optum\textsuperscript{TM} EHR data.  This data set is comprised of EHR records from hospitals, hospital networks, general practice offices and specialist clinical providers across the United States of America.  It includes anonymized patient demographics, hospitalizations, laboratory tests and results, in-patient and prescribed medications, procedures, observations and diagnoses.  The data provides most information collected during a patient journey provided all care sites for a given patient are included in the list of Optum\textsuperscript{TM} data contributors.  This data set is one of the most comprehensive EHR databases in the world\footnote{https://www.optum.com/business/life-sciences/real-world-data/ehr-data.html} and is used extensively for real-world clinical studies \cite{dagenais2022use}.

\begin{figure}[H]
\centering
\begin{tabular}{cc}
  \includegraphics[width=0.8\linewidth]{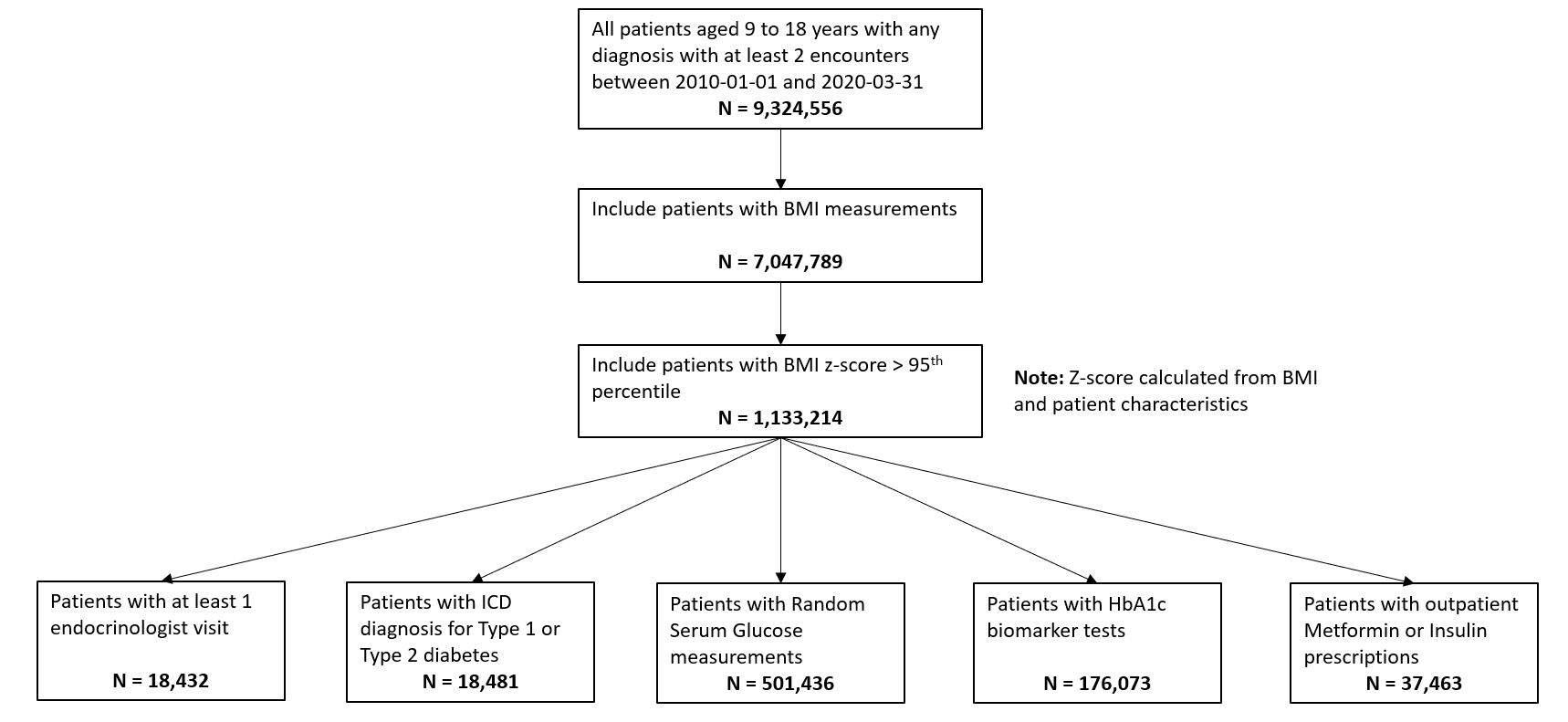}
\end{tabular}
\caption{Data specification for pediatric patients at risk of T2DM in Optum\textsuperscript{TM} EHR database. The BMI Z-scores were calculated using Equation 1 from the CDC Growth Charts.  The bottom row shows the number of patients having the specified study characteristics for the pediatric T2DM phenotype for the relevant variables.}
\label{fig:attrition}
\end{figure}

\vspace{5mm}
We could not use the same data set from Hubbard et al. as that is private to the PEDSnet consortium \footnote{https://pedsnet.org/about/our-story/}.  Instead, we extracted a patient cohort of pediatric patients with equivalent T2DM risk characteristics as defined by Hubbard et al. for this analysis.  The Pediatric T2DM data structure and Bayesian latent phenotype model were reproduced in order to test how well the proposed Bayesian LCA model translates from the PEDSnet EHR data to the Optum\textsuperscript{TM} EHR data. Hubbard et al. used data from a single site, the Children’s Hospital of Philadelphia (N = 68,265). The Optum data covered all regions of the United States of America (N = 1,133,215). The PEDSnet data used in the Hubbard et al. work are located within the US Northeast region which comprises about 13\% of the Optum data.  To account for potential variance of pediatric T2DM prevalence across the USA we used the Northeast subset of the Optum\textsuperscript{TM} EHR for the Bayes LCA comparison, although we also ran the study using all of the data.  The data specification in Figure ~\ref{fig:attrition} shows how the data was extracted from Optum\textsuperscript{TM} EHR.  The overall objective for this specification was to arrive at the same patient identification rules used by Hubbard et al.  We also restricted the data variables to those used in the Hubbard et al. study.  This resulted in a large proportion of duplicate observations when the unique patient ID is removed prior to running the model as most variables are binary indicators, the BMI Z-score is restricted to high percentiles and the two biomarker variables therefore have a narrow range of values (plus many missing values).  This decreased the unique data size to approximately 435,000 unique observations which remains a very large data set.  The BMI Z-score was calculated using Equation 1 from the CDC Growth Charts \footnote{https://www.cdc.gov/nccdphp/dnpa/growthcharts/resources/biv-cutoffs.pdf}

\begin{figure}[H]
  \captionof{table}{Comparison of PEDSnet (A) and Optum\textsuperscript{TM} (B) study population characteristics}
  \centering
  \includegraphics[height=7cm]{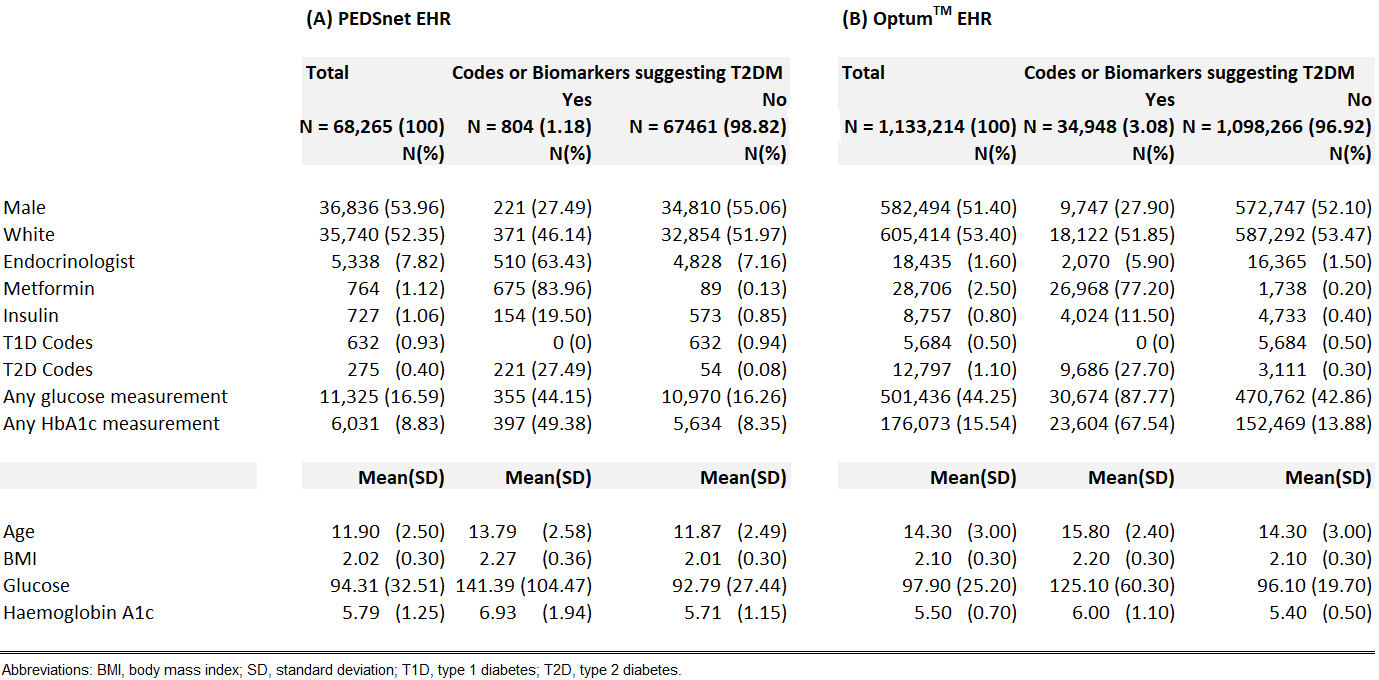}
  \label{table:data}
\end{figure}

The full USA Optum\textsuperscript{TM} EHR pediatric T2DM data extraction is compared with Hubbard et al. PEDSnet data in Table ~\ref{table:data}.  The majority of proportions are similar between the two data sets. Notable differences are endocrinologist visits (substantially fewer in Optum\textsuperscript{TM} EHR) and availability of biomarker measurements (substantially more in Optum\textsuperscript{TM} EHR).  The Northeast subset had closer proportions.

\section{Methods}

Our aim is to investigate if the Bayes LCA model proposed by Hubbard et al. \cite{hubbard2019bayesian} can be generalised and expanded to other data and disease conditions.  We first reproduced the model as defined with JAGS MCMC using the Optum\textsuperscript{TM} EHR T2DM data.  This gives us confidence that the model can generalise to other EHR data.  We then followed up with the \textit{Stan} \cite{carpenter2017stan} implementations of Hamiltonian MCMC and VB to investigate how the model specification translates to other algorithmic approaches.  We are particularly interested in how well the model scales to realistically large real-world clinical data, so we have focused in particular on variational Bayes algorithms.

\subsection{Bayes LCA Model}

The canonical form for LCA assumes the response $\textbf{y}$ is categorical and each observation $i$ with observed variables $J = \{1,...,J\}$ belongs to one of $\textbf{C}$ classes, represented by a latent class indicator $z_i = \{1,...,C\}$.  The marginal response probabilities are

\vspace{5mm}
\begin{center}
$P(\textbf{y}_i) = \sum\limits_{c=1}^C\pi_c\prod\limits_{j=1}^JP(y_{ij}|z_i=c) \nonumber$
\end{center}
\vspace{5mm}

Where $\textbf{y}_i$ the response vector $(y_{i1},...,y_{iJ})^T$ for observation $i$ on $\textbf{J}$ variables, $\textbf{z}_i$ is the latent class that observation $i$ belongs to and $\boldsymbol{\pi}_c$ is the probability of being in class $c$.

The Bayes LCA model applied is from Hubbard et al. \cite{hubbard2019bayesian} and follows the general specification shown in Table ~\ref{table:modelSpec}.  In the Hubbard et al. model there are two classes in $\textbf{C}$ (has T2DM or does not have T2DM) and three sets of two items each (2 biomarker indicators, 2 ICD disease codes and 2 diabetes medications).  The latent phenotype variable for each patient, $D_i$, is assumed to be associated with patient characteristics that include demographics, biomarkers, clinical diagnostic codes and prescription medications.  This model allows for any number of clinical codes or medications and in this model each clinical code and medication are binary indicator variables to specify if the code or medication was present for that patient.  Since it is common for specific biomarker laboratory tests to be missing across a cohort of patients, the model includes an indicator variable per biomarker to indicate whether or not specific biomarker data is available for that patient.  Biomarker availability is an important component of the model since biomarkers are widely considered to be high quality prognostic data for many disease areas \cite{hadjadj2004prognostic}.

\begin{center}
\captionof{table}{Model specification for Bayesian latent variable model for EHR-derived phenotypes for patient $i$.}
\vspace{5mm}
\scalebox{0.80}{
\begin{tabular}{lllll}
                                        & \multicolumn{1}{c}{\bfseries Variable} & \multicolumn{1}{c}{\bfseries Model} & \multicolumn{1}{c}{\bfseries Priors}   \\
\toprule
                                        &                       &       &            \\
\textbf{Latent Phenotype}                        & $D_i$                 & $D_i \sim $Bern$(g(\bm{X}_i\bm{\beta}^D + \eta_i))$            & $\beta^D \sim $MVN$(0, \Sigma_D); \eta_i \sim $Unif$(a,b)$              \\
                                        &                       &       &            \\
\textbf{Availability of Biomarkers}              & $R_{ij}, j=1,...,J$   & $R_{ij} \sim $Bern$(g((1,\bm{X}_i,D_i)\bm{\beta}^R_j))$        & $\beta^R_j \sim $MVN$(\mu_R, \Sigma_R)$                                  \\
                                        &                       &       &            \\
\textbf{Biomarkers}                              & $Y_{ij}, j=1,...,J$   & $Y_{ij} \sim $N$(g((1,\bm{X}_i,D_i)\bm{\beta}^Y_j, \tau^2_j))$ & $\beta^Y_j \sim $MVN$(\mu_Y, \Sigma_Y); \tau^2_j \sim $InvGamma$(c,d)$  \\
                                        &                       &       &            \\
\textbf{Clinical Codes}                          & $W_{ik}, k=1,...,K$   & $W_{ik} \sim $Bern$(g((1,\bm{X}_i,D_i)\bm{\beta}^W_k))$        & $\beta^W_k \sim $MVN$(\mu_W, \Sigma_W)$                                  \\
                                        &                       &       &            \\
\textbf{Prescription Medications}                & $P_{il}, l=1,...,L$   & $P_{il} \sim $Bern$(g((1,\bm{X}_i,D_i)\bm{\beta}^P_l))$        & $\beta^P_l \sim $MVN$(\mu_P, \Sigma_P)$                                  \\

                                        &                       &       &            \\
\bottomrule
                                        &                       &       &            \\
                                        &                       &       & \multicolumn{1}{c}{$\bm{ g(\bm{\cdot}) = exp(\bm{\cdot})/(1 + exp(\bm{\cdot}))}$} \\
\label{table:modelSpec}
\end{tabular}}
\end{center}


The model likelihood for the $i^{th}$ patient is given by

\begin{align}
\mathcal{L}(\eta_i,\beta^D,\beta^R,\beta^Y,\beta^W,\beta^P,\tau^2|X_i) & = \sum_{d=0,1}{P(D_i=d|\eta_i,\beta^D,X_i)} \nonumber \\
& \prod_{j=1}^{J}{f(R_{ij}|D_i=d,X_i,\beta_j^R)f(Y_{ij}|D_i=d,X_i,\beta_j^Y,\tau_j^2)^{R_{ij}}} \nonumber \\
& \prod_{k=1}^{K}{f(W_{ik}|D_i=d,X_i,\beta_k^W)} \prod_{l=1}^{L}{f(P_{il}|D_i=d,X_i,\beta_i^P)} \\ \nonumber
\end{align}

where $\beta^D$ associates the latent phenotype to patient characteristics, $\eta_i$ is a patient-specific random effect parameter and parameters $\beta^R,\beta^Y,\beta^W$ and $\beta^P$ associate the latent phenotype to patient characteristics and biomarker availability, biomarker values, clinical codes and medications respectively.  $X_i$ represents $M$ patient covariates such as demographics ($X_i = X_{i1},...,X_{i,M}$).  The mean biomarker values are shifted by a regression quantity $\beta^Y_{j,M+1}$ for patients with the phenotype compared to those without. Sensitivity and specificity of binary indicators for clinical codes, medications and presence of biomarkers are given by combinations of regression parameters. For instance, in a model with no patient covariates, sensitivity of the $k^{th}$ clinical code is given by $expit(\beta^W_{k_0} + \beta^W_{k_1})$, while specificity is given by $1-expit(\beta^W_{k_0})$, where $expit(\cdot) = exp(\cdot)/(1+exp(\cdot))$.

\vspace{5mm}
We validated this model with a real-world example, namely pediatric T2DM.  Table ~\ref{table:modelExample} indicates how the Bayes LCA phenotyping model maps to this disease area.

\begin{center}
\captionof{table}{Mapping the example pediatric T2DM study for patient $i$ as defined in Hubbard et al. \cite{hubbard2019bayesian} to the Bayes LCA phenotyping model factors. }
\vspace{5mm}
\scalebox{0.80}{
\begin{tabular}{lllll}
                                        & \multicolumn{1}{c}{\bfseries Model Variable} & \multicolumn{1}{c}{\bfseries Data Elements}   \\
\toprule
                                        &                       &                  \\
\textbf{Latent Phenotype}               & $D_i$                 & Type 2 Diabetes (latent variable not contained within data set)   \\
                                        &                       &                   \\
\textbf{Availability of Biomarkers}     & $R_{ij}, j=1,...,J$   & Availability of ($j$=1) Glucose and ($j$=2) HbA1c biomarker data  \\
                                        &                       &                  \\
\textbf{Biomarkers}                     & $Y_{ij}, j=1,...,J$   & Laboratory test values for ($j$=1) Glucose and ($j$=2) HbA1c  CPT codes\\
                                        &                       &                   \\
\textbf{Clinical Codes}                 & $W_{ik}, k=1,...,K$   & ($k$=1) ICD Code for T2 Diabetes and ($k$=2) CPT code for Endocrinologist Visit\\
                                        &                       &                   \\
\textbf{Prescription Medications}       & $P_{il}, l=1,...,L$   & NCD Medication codes for ($l$=1) Insulin and ($l$=2) Metformin  \\

                                        &                       &                   \\
\bottomrule
\label{table:modelExample}
\end{tabular}}
\end{center}

\vspace{5mm}
Although the pediatric T2DM example has two levels for $J, K$ and $L$, the model can be expanded to multiple biomarkers, clinical codes and medications for other disease areas.  The model easily extends to include additional elements.  For example, the number of hospital visits or the number of medication prescriptions, or, for other disease conditions or study outcomes, comorbidities and medical costs etc. We extended the Hubbard et al. work by comparing a range of alternative methods to \textit{JAGS} that included alternative MCMC via \textit{Stan} \cite{carpenter2017stan} as well as VB (also using \textit{Stan}). Our objectives were to understand the advantages and challenges posed by the Bayes LCA model using these approaches compared to the traditional sampling MCMC approach.  The following sub-sections discuss each of these approaches in turn.

\subsection{Gibbs Monte-carlo Sampling}

We used \textit{JAGS} \cite{plummer2003jags} as the baseline approach with which to compare all other methods since it uses the most parsimonious method (Gibbs/Metropolis-Hastings) \cite{geman1984stochastic}.  Our initial objective was to reproduce the Hubbard et al. model using the same methods they employed but against a different EHR database.  We used the same JAGS LCA model published by Hubbard et al. in their GitHub supplement \footnote{https://github.com/rhubb/Latent-phenotype/}.  The results obtained were close to the Hubbard et al. results with the Northeast region subset of Optum\textsuperscript{TM} (Table ~\ref{table:comp}).  The differences in biomarker shift can be explained by the wider patient geographic catchment and differences in missing data for PEDSnet and Optum\textsuperscript{TM}. Following these results we are confident in the extensibility of the model to the same disease area with different data.  Unfortunately the computational performance is poor using \textit{JAGS} with $\sim$38,000 observations, taking 16 hours on an Intel core i9 computer with 64GB memory.

\vspace{10cm}
\begin{center}
\captionof{table}{Comparison of Optum\textsuperscript{TM} results with Hubbard et al. PEDSnet using the same JAGS LCA model published in Hubbard et al. Github\cite{hubbard2019bayesian}}
\vspace{5mm}
\scalebox{0.90}{
\begin{tabular}{llll}
\textbf{}                               & \multicolumn{2}{l}{\textbf{Posterior Mean (95\% CI)}}                   \\
                                        & (a) PEDSnet data                   & (b) Optum\textsuperscript{TM} data \\
\toprule
                                        &                                    &                                    \\
T2DM code sensitivity (expit($\beta^W_{10} + \beta^W_{11}$))                   & 0.17 (0.15, 0.20)       &  0.15 (0.12, 0.18)      \\[1mm]
T2DM code specificity (1-expit($\beta^W_{10}$))                   & 1.00 (1.00, 1.00)       &  1.00 (1.00, 1.00)                   \\[1mm]
Endocrinologist visit code sensitivity (expit($\beta^W_{20} + \beta^W_{21}$))  & 0.94 (0.92, 0.95)       &  0.18 (0.15, 0.21)      \\[1mm]
Endocrinologist visit code specificity (1-expit($\beta^W_{20}$))  & 0.93 (0.93, 0.94)       &  0.99 (0.98, 0.99)                   \\[1mm]
Metformin code sensitivity (expit($\beta^P_{10} + \beta^P_{11}$))              & 0.31 (0.28, 0.35)       &  0.40 (0.36, 0.44)      \\[1mm]
Metformin code specificity (1-expit($\beta^P_{10}$))              & 0.99 (0.99, 0.99)       &  0.98 (0.98, 0.99)                   \\[1mm]
Insulin code sensitivity (expit($\beta^P_{20} + \beta^P_{21}$))                & 0.66 (0.61, 0.70)       &  0.55 (0.51, 0.59)      \\[1mm]
Insulin code specificity (1-expit($\beta^P_{20}$))                & 1.00 (1.00, 1.00)       &  1.00 (1.00, 1.00)                   \\[1mm]
Mean shift in HbA1c ($\beta^Y_{12}$)                     & 3.15 (3.06, 3.24)       &  4.80 (4.72, 4.81)                            \\[1mm]
Mean shift in glucose ($\beta^Y_{11}$)                   &            90.62 (90.25, 91.00)    &            89.30 (89.10, 90.01)    \\[1mm]
\bottomrule
\label{table:comp}
\end{tabular}}
\end{center}

\subsection{Hamiltonian Monte-Carlo Sampling}

We used \textit{Stan} MC \cite{carpenter2017stan} for an alternative MCMC comparison.  Hamiltonian Monte-Carlo (and the default extension used by \textit{Stan} MC called 'No U-turn Sampler (NUTS)') includes a gradient optimisation step so it cannot sample discrete latent parameters in the way JAGS can.  Instead, \textit{Stan} MC integrates the posterior distribution over the discrete classes \cite{yackulic2020need} so this is a useful comparison to the fully sampling Gibbs approach.  We translated the \textit{JAGS} model directly to a \textit{Stan} model using similar sampling notation.  Although this has reasonable results (table ~\ref{table:allResults}) we feel it may benefit from reparameterization \cite{dombowsky2020assessing} and/or use of \textit{Stan}-specific helper functions e.g. log-sum-exp and the target log-probability increment statement approach to the model definition.  In particular, reparameterization may improve the variational inference results, which are a struggle to tune effectively.

\subsection{Variational Inference}

\begin{figure}[H]
\centering
  \includegraphics[height=8cm]{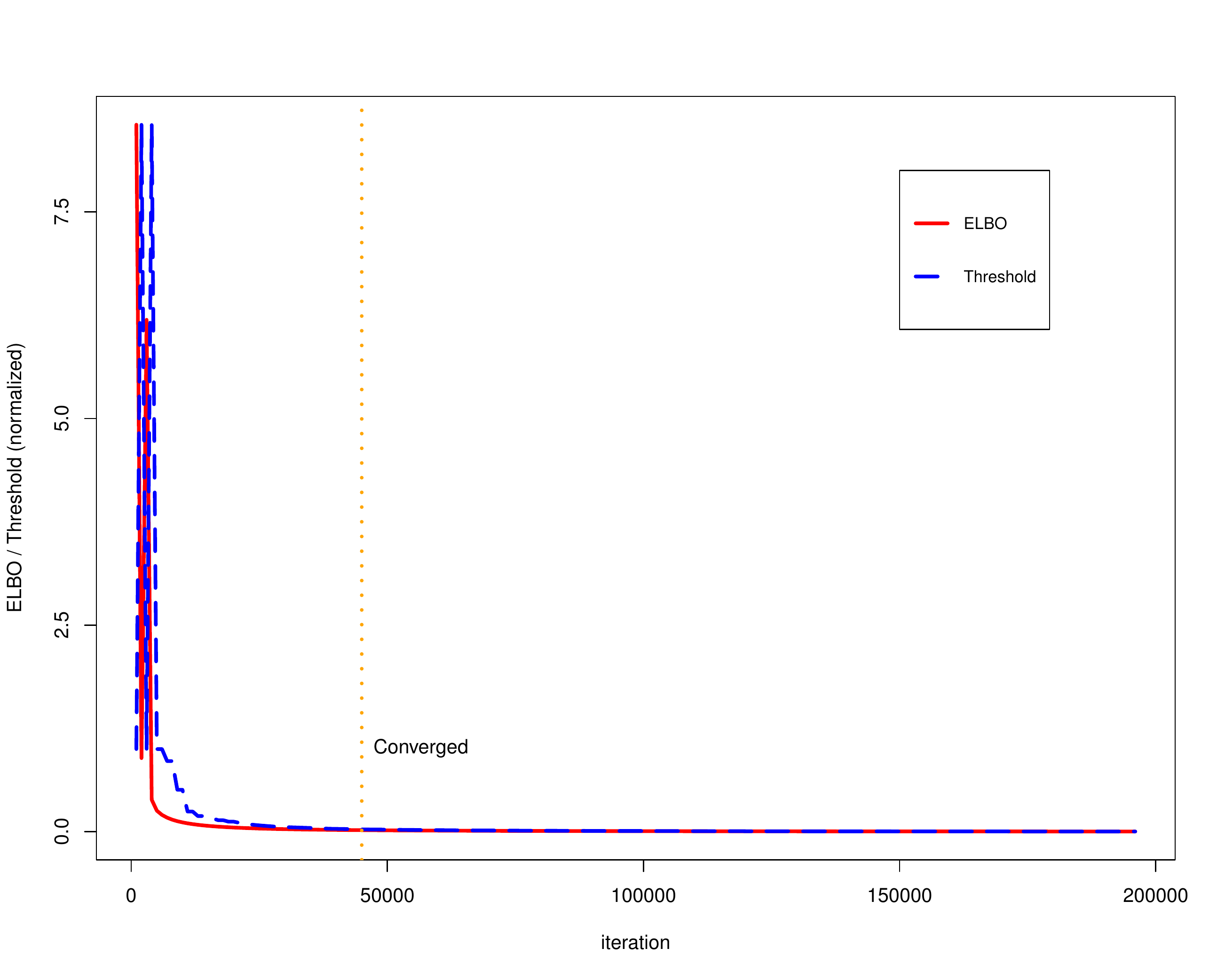} \\
\caption{Runtime ELBO (solid) and threshold delta (dashed) for all iterations applied.  ELBO and threshold have been normalized to the same scale.  We can see there is a diminishing return after about 45,000 iterations (vertical dotted line).  In this example the algorithm ran for over a day longer than it needed to in finding the best posterior estimate it could generate.}
\label{fig:elboplot}
\end{figure}

We used \textit{Stan} VB \cite{kucukelbir2017automatic} for variational Bayes.  Our previous baseline work with a logistic regression model using Pima Indian data \footnote{see supplementary material for details of the Pima Indian baseline study} highlighted the challenges in using automatic variational methods (i.e. not using a closed-form solution for the evidence lower bound optimization).  In \textit{Stan} there is only one implementation of variational inference, the automatic differentiation approach \cite{kucukelbir2015automatic}.  We found this worked well for the simple Pima Indian logistic regression model but is more challenging for the Bayes LCA model.  For both \textit{Stan} HMC and VB, we used the same \textit{Stan} model definition\footnote{see supplementary material for the \textit{Stan} LCA model details}.  Reparameterization of the LCA model and use of \textit{Stan}-specific helper functions as discussed above may improve the predictive performance for VB.  There are currently no closed-form solutions implemented in R or Python for variational Bayes LCA.  Given the Pima Indian results demonstrated the effectiveness of closed-form conditionally conjugate solutions for logistic models we suggest this is an interesting topic for further work on variational Bayes LCA.  Figure ~\ref{fig:elboplot} illustrates a common problem with VB.  There are two hyperparameters for stopping, a maximum number of ELBO calculation iterations and a difference threshold (delta) between iteration $t$ and $t+1$.  In Figure ~\ref{fig:elboplot} we can see that the $argmin \{ELBO\}$ has effectively converged after approximately 45,000 iterations but we have set maximum iterations to 200,000 and the threshold tolerance relatively low meaning it is never reached so the algorithm continues well past the best estimate it can produce.  Unfortunately, there is no way \textit{a priori} to know what a suitable threshold tolerance should be as ELBO values are unbounded, or the effective number of iterations for ELBO convergence as this depends on various factors including model specification and algorithm tuning.  We therefore risk running the VB algorithm much longer than needed to reach the best posterior estimate it can deliver under a specific model.  However the variational approach consistently uses significantly less computer memory than MCMC.  With this EHR phenotyping model on our computer system VB used 1.8GB versus 37GB of RAM memory for a three-chain MCMC
.

\subsection{Maximum Likelihood Approach}

We used the R package, \textit{clustMD} \cite{mcparland2016model} to compare a maximum likelihood (MLE) clustering approach to Bayesian LCA.  \textit{clustMD} employs a mixture of Gaussian distributions to model the latent variable and an Expectation Maximisation (EM) algorithm to estimate the latent cluster means.  It also employs Monte Carlo EM for categorical data.  \textit{clustMD} supports mixed data so is appropriate in our context.  To use \textit{clustMD} the data must be reordered to have continuous variables first, followed by ordinal variables and then nominal variables.  For our data the computational runtime was about 50\% of \textit{JAGS} MCMC, approximately 62 hours.


\section{Results}

We tested posterior diagnostics, goodness of fit diagnostics and the empirical performance of the LCA model.  \textit{Stan} has comprehensive posterior diagnostics available via the \textit{posterior} and \textit{bayesplot} R packages.  The \textit{loo} R package provides goodness of fit diagnostics based on Pareto Smoothed Importance Sampling (PSIS) \cite{vehtari2015pareto}, leave-one-out cross-validation and the Watanabe-Akaike/Widely Applicable information criterion (WAIC) \cite{magnusson2020leave} and \cite{vehtari2017practical}.

\begin{center}
\captionof{table}{Comparison of LCA model results for clinical attributes}
\vspace{3mm}
\scalebox{0.80}{
\begin{tabular}{lllll}
\textbf{}                               & \multicolumn{3}{l}{\textbf{Posterior Mean (95\% CI)}}                                                     \\
                                        & (a) JAGS Gibbs MCMC           & (b) Stan HMC           & (c) Stan VB                                      \\
\toprule
                                        &                         &                       &                                                         \\
T2DM code sensitivity (expit($\beta^W_{10} + \beta^W_{11}$))                   &  0.15 (0.12, 0.18)      &  0.10 (0.09, 0.11)    & 0.12 (0.10, 0.12)      \\[1mm]
T2DM code specificity (1-expit($\beta^W_{10}$))                     &  1.00 (1.00, 1.00)      &  1.00 (0.99, 1.00)    & 0.99 (0.99, 0.99)      \\[1mm]
Endocrinologist visit code sensitivity (expit($\beta^W_{20} + \beta^W_{21}$))   &  0.18 (0.15, 0.21)      &  0.20 (0.18, 0.21)    & 0.22 (0.19, 0.22)      \\[1mm]
Endocrinologist visit code specificity (1-expit($\beta^W_{20}$))  &  0.99 (0.98, 0.99)      &  0.98 (0.97, 0.99)    & 0.97 (0.97, 0.99)      \\[1mm]
Metformin code sensitivity (expit($\beta^P_{10} + \beta^P_{11}$))              &  0.40 (0.36, 0.44)      &  0.21 (0.20, 0.21)    & 0.19 (0.19, 0.20)      \\[1mm]
Metformin code specificity (1-expit($\beta^P_{10}$))               &  0.98 (0.98, 0.99)      &  0.93 (0.92, 0.93)    & 0.93 (0.92, 0.94)      \\[1mm]
Insulin code sensitivity (expit($\beta^P_{20} + \beta^P_{21}$))                &  0.55 (0.51, 0.59)      &  0.35 (0.31, 0.35)    & 0.20 (0.19, 0.20)      \\[1mm]
Insulin code specificity (1-expit($\beta^P_{20}$))                &  1.00 (1.00, 1.00)      &  1.00 (0.99, 1.00)    & 1.00 (0.99, 1.00)      \\[1mm]
Mean shift in HbA1c ($\beta^Y_{12}$)                     &  4.80 (4.72, 4.81)      &  4.77 (4.76, 4.78)    & 4.77 (4.75, 4.78)      \\[1mm]
Mean shift in glucose ($\beta^Y_{11}$)                   & 89.30 (89.10, 90.01)                                     & 88.59 (88.48, 88.71)                                   & 22.80 (21.06, 24.92) \\
\bottomrule
\label{table:allResults}
\end{tabular}}
\end{center}

\subsection{Posterior Diagnostics}

\begin{figure}[H]
\centering
\scalebox{0.90}{
\begin{tabular}{cc}
  \includegraphics[width=0.45\linewidth]{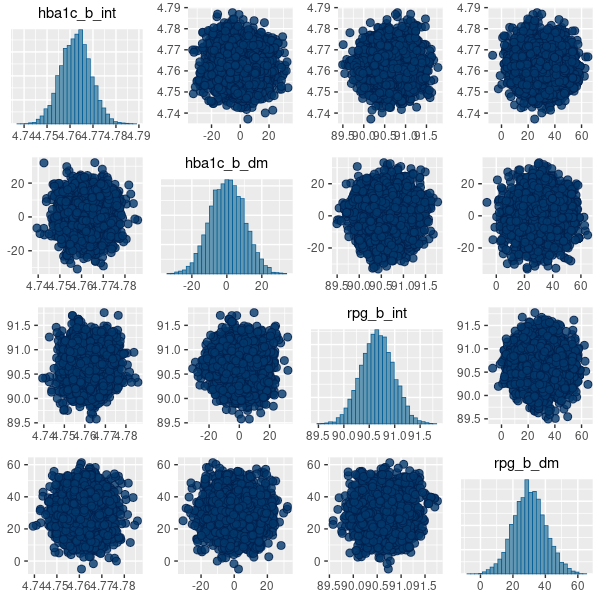} & \includegraphics[width=0.45\linewidth]{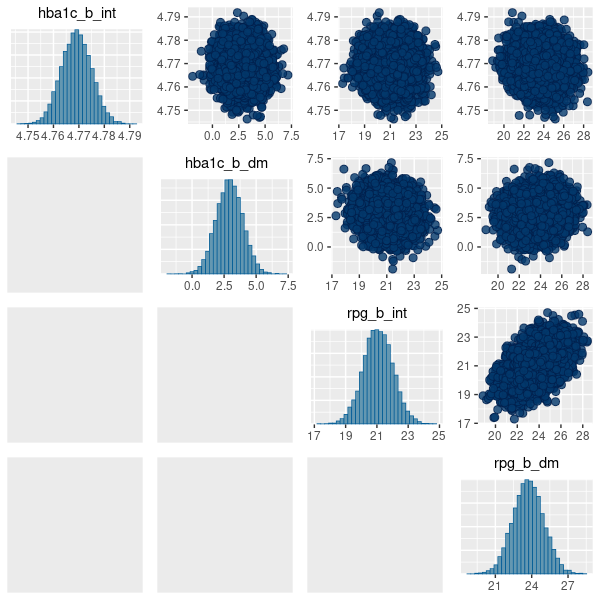} \\
  \fontsize{18}{18}\selectfont{\textbf{(a)}} & \fontsize{18}{18}\selectfont{\textbf{(b)}} \\
\end{tabular}}
\caption{\textit{bayesplot} pairs plots for \textit{Stan} HMC (a) and \textit{Stan} VB (b) for the two biomarkers.  Each biomarker contains two model components as defined in the model specification (Table ~\ref{table:modelSpec}).  The b\_int component is $N(X_i\beta^Y_j,\tau^2_j)$ and the b\_dm component is $N(D_i\beta^Y_j,\tau^2_j)$ as described in Hubbard et al. Section 2.2 \cite{hubbard2019bayesian}.}
\label{fig:bayesplot}
\end{figure}

The posterior diagnostics plots for the biomarkers in Figure ~\ref{fig:bayesplot} show that, for MCMC (a), we see no evidence of collinearity and the posterior means appear close to those we obtained with \textit{JAGS}.  The \textit{mcmc pairs} plot for VB (b) appears reasonable for HbA1c but not for the Glucose biomarker \textit{rpg\_b\_int} component.  There appears to be correlation between the glucose components.  Since there is a single 'chain' for VB there is only the top triangular set which represents 100\% of the posterior samples.  This type of plot does not communicate the relative variances of the posteriors.

\subsection{Goodness of fit}

\begin{figure}[H]
\centering
\scalebox{0.90}{
\begin{tabular}{cc}
  \includegraphics[width=0.5\linewidth]{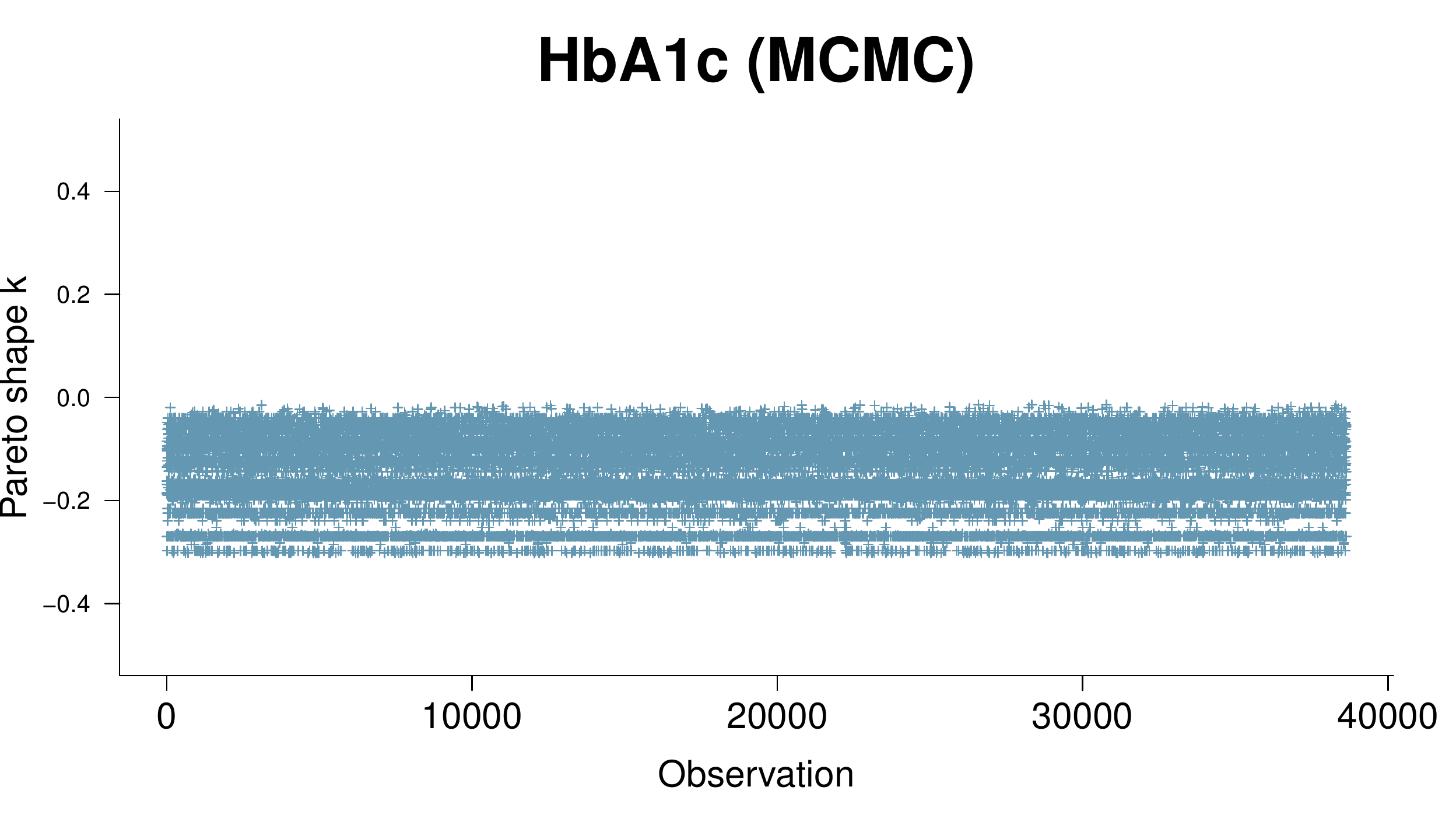} & \includegraphics[width=0.5\linewidth]{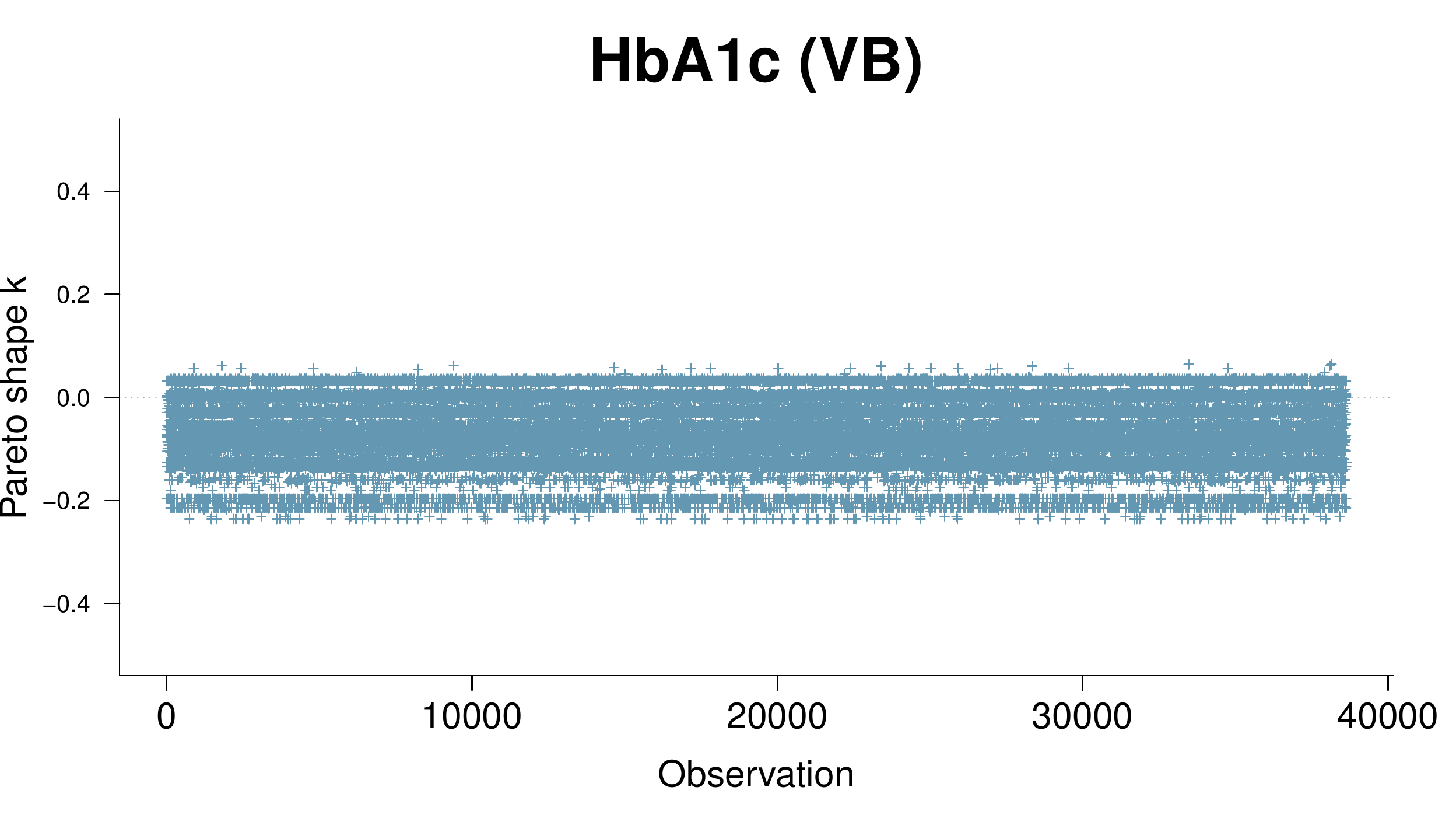} \\
  & \multicolumn{1}{l}{\fontsize{16}{16}\selectfont{\textbf{(a)}}}  \\
  \includegraphics[width=0.5\linewidth]{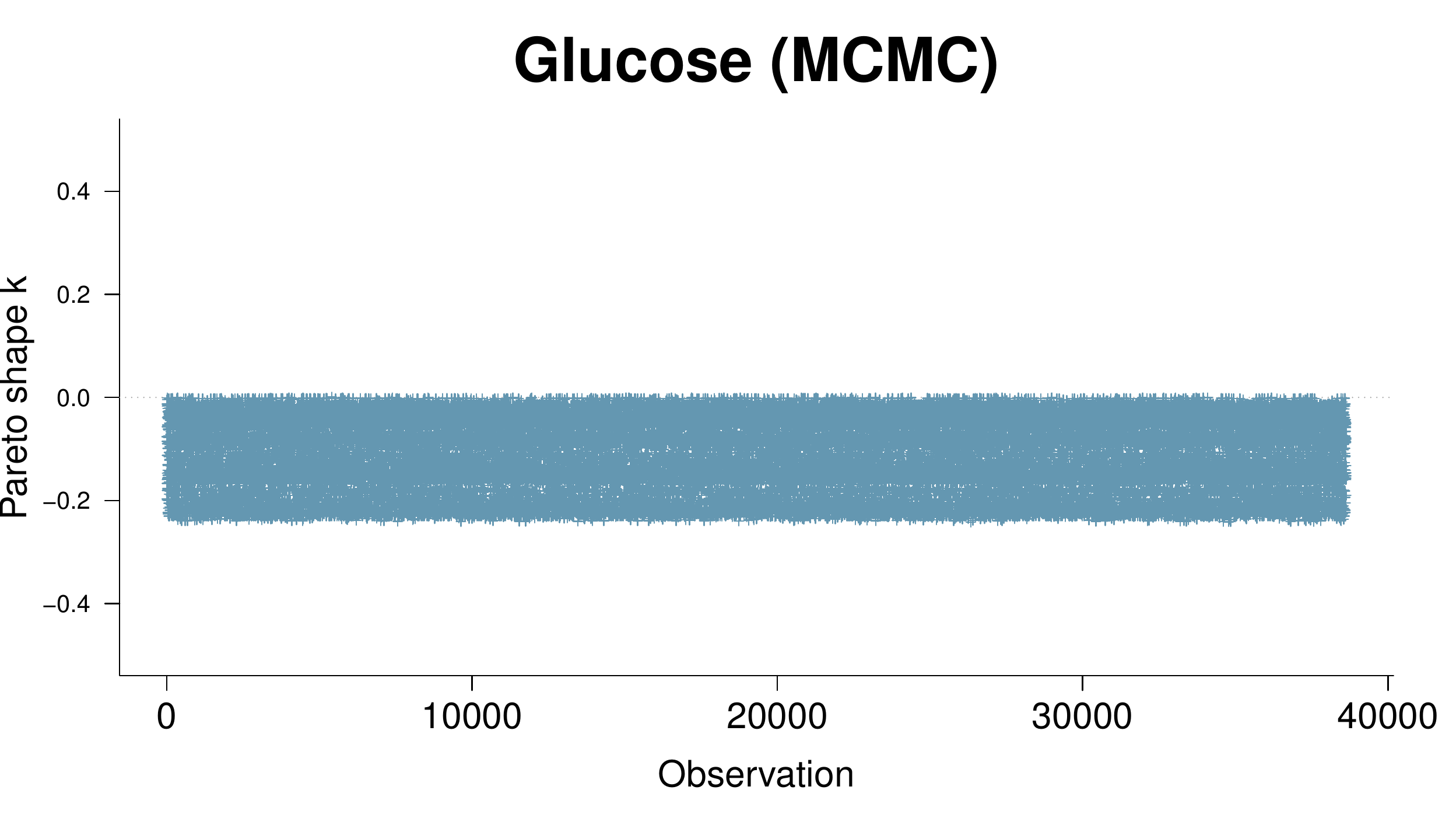} & \includegraphics[width=0.5\linewidth]{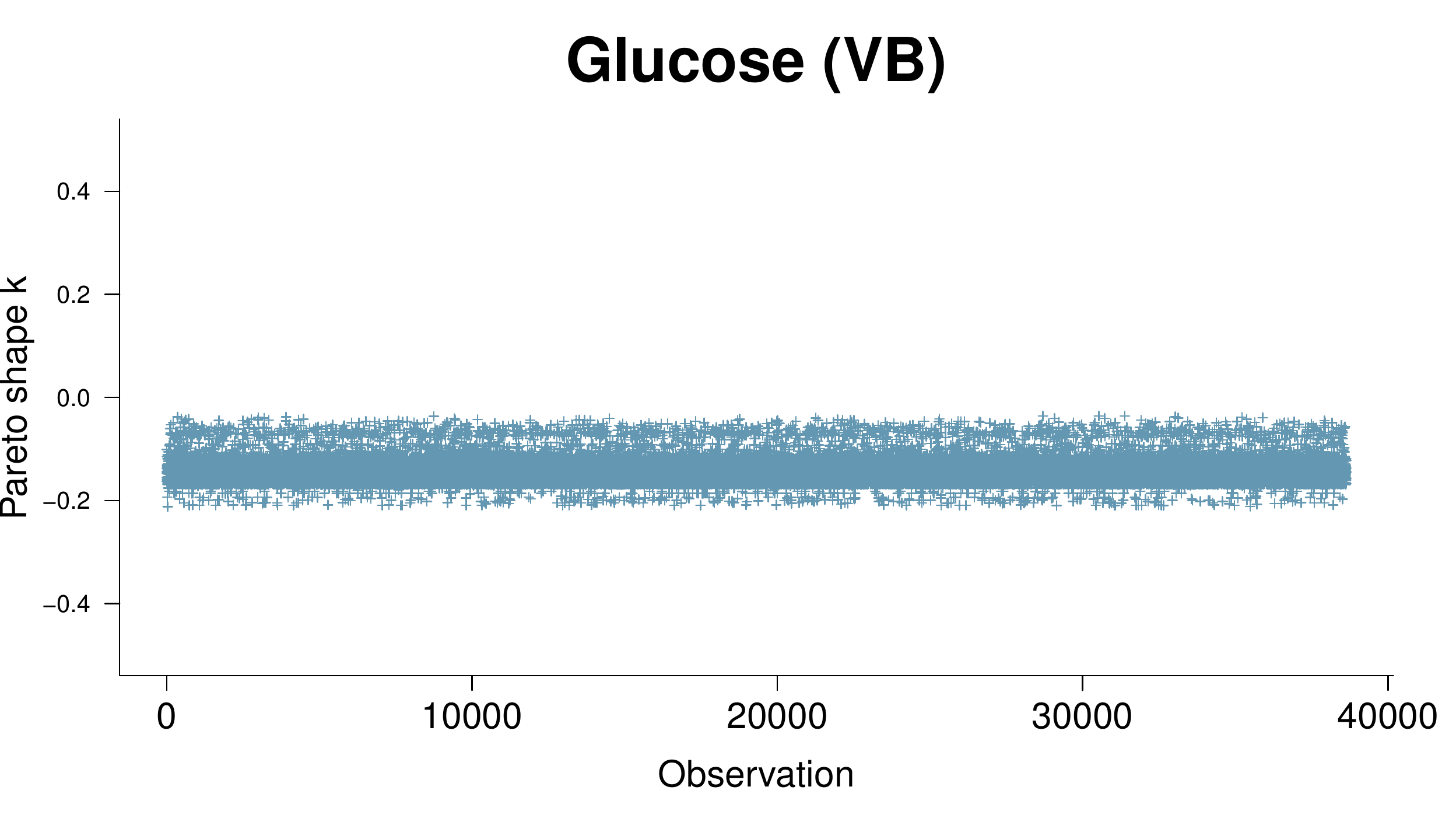} \\
  & \multicolumn{1}{l}{\fontsize{16}{16}\selectfont{\textbf{(b)}}} \\
\end{tabular}}
  \caption{PSIS plots for the two biomarkers, HbA1c (a) and Random Glucose (b) under MCMC (left) and VB (right). Both biomarkers are well below $k$=0.5.  The HbA1c biomarker is slightly worse for VB with a segment of observations above 0 but is well within good territory.  The Random Glucose biomarker appears better in VB compared to MCMC but we know that the expected value obtained by VB for Glucose is not close to the true value obtained by MCMC.}
  \label{fig:loo}
\end{figure}

We used approximate leave-one-out cross-validation from the R \textit{loo} package to evaluate the goodness of fit for the model.  \textit{loo} uses log-likelihood point estimates from the model to measure its predictive accuracy against training samples generated by Pareto Smoothed Importance Sampling (PSIS) \cite{vehtari2015pareto}.  The PSIS shape parameter $k$ is used to assess the reliability of the estimate.  If $k$ < 0.5 the variance of the importance ratios is finite and the central limit theorem holds.  If $k$ is between 0.5 and 1 the variance of the importance ratios is infinite but the mean exists.  If $k$ > 1 the variance and mean of the importance ratios do not exist. The results for the two biomarkers (Figure ~\ref{fig:loo}) show all of the n=38,000 sample are good with no outliers.  It appears VB performs approximately as well as MCMC for this metric.

\subsection{Empirical Performance}

The model mean estimates for biomarkers and the sensitivity analysis for the indicator variables show good agreement with \textit{JAGS} apart from the Glucose variable under ADVI (Table 4). The VB results may be improved after reparameterization of the model and use of \textit{Stan}-specific helper functions\footnote{See "Reparameterization and Change of Variables" section in the \textit{Stan} user manual}. 

\subsection{Maximum Likelihood Approach}

\begin{figure}[H]
\centering
\begin{tabular}{cc}
  \includegraphics[width=0.8\linewidth,height=6cm]{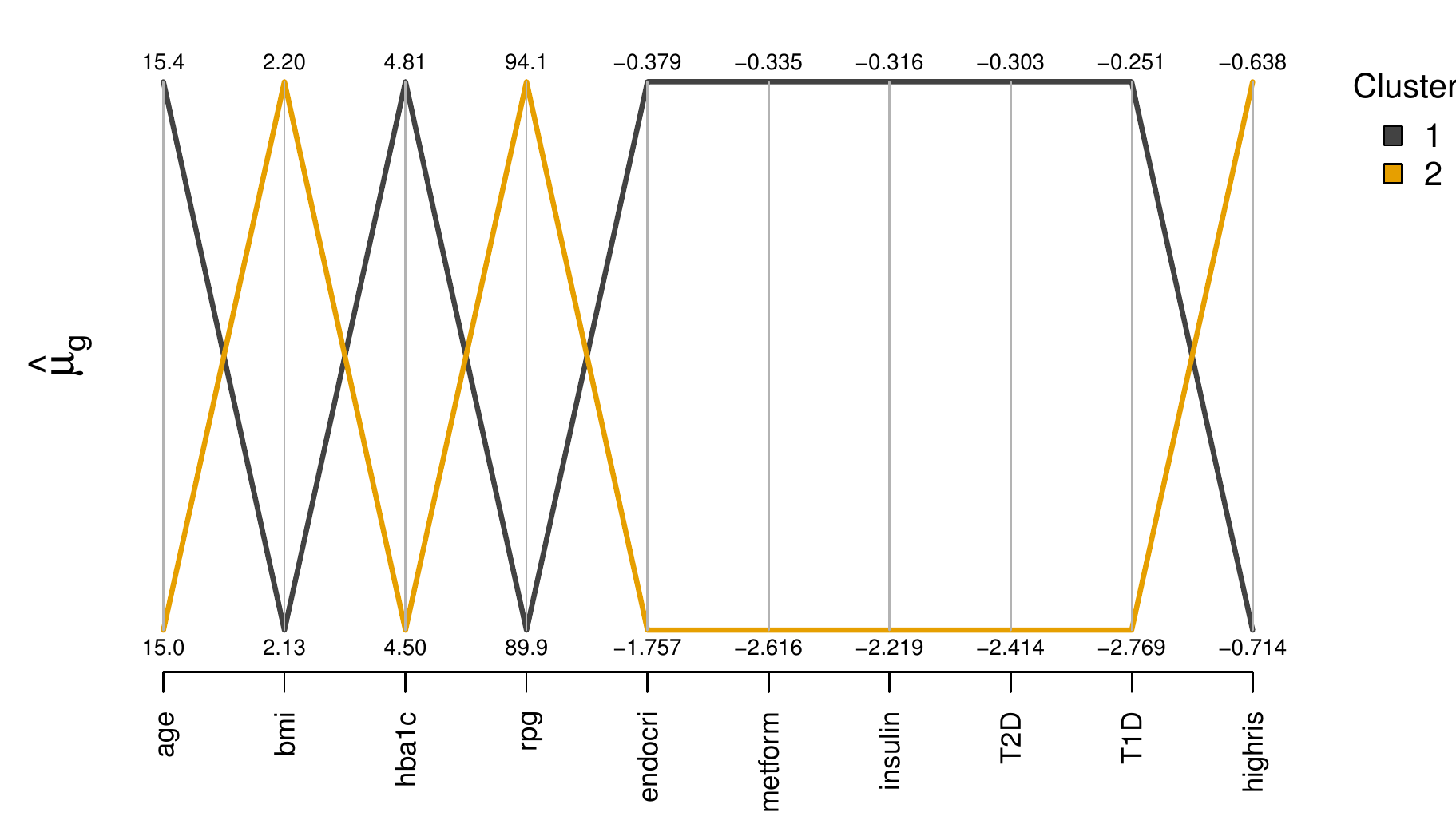} \\
  \includegraphics[width=0.8\linewidth,height=6cm]{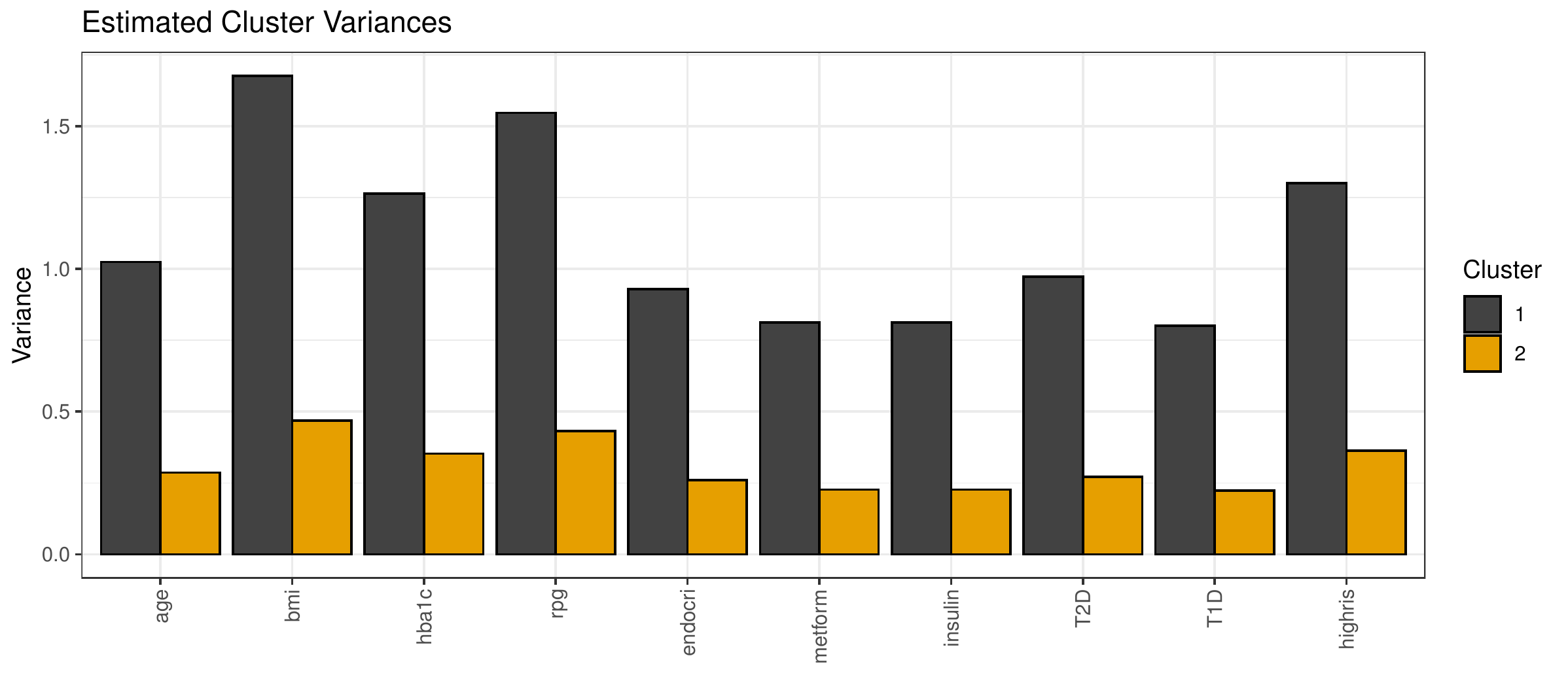}
\end{tabular}
\caption{\textit{clustMD} plots running 2 latent clusters.}
\label{fig:mle}
\end{figure}

For \textit{clustMD} we take cluster 1 as the T2DM class as it is the minority cluster.  The cluster means parallel coordinates plot in Figure ~\ref{fig:mle}(a) indicates similar HbA1c (4.81) and Glucose (89.9) levels compared to the Bayesian model. Cluster 1 has a significantly larger variance (Figure ~\ref{fig:mle}(b)) that might be due to the high imbalance of T2DM in the data.  Comparison between this MLE approach and the Bayesian methods for the discrete variables is not clear as we are focusing on the posterior priors in our Bayes model to determine sensitivity.

\section{Discussion}

We have demonstrated that this factorised form of Bayes LCA model generalises to other EHR data. We believe this model form should also generalise to other disease areas.  It can potentially become a uniquely useful tool for clinical decision support, especially for rare disease areas where gold-standard data is sparse. We have compared a number of alternative implementations of the model to identify if the intrinsic computation limitations from MCMC can be overcome in a real-world setting using VB. We have compared a range of methods for similar problems and set out practical guidance on implementing such models. We have investigated the challenges and suggest potential solutions for each of the alternative methods studied.  For LCA, there are no closed-form implementations currently available so we need to use automatic "black box" approaches.  We find automatic VB methods as implemented both for the baseline logistic model and in \textit{Stan} VB are complex to configure and are very sensitive to model definition, algorithm hyperparameters and choice of gradient optimiser. The LCA model was significantly more challenging to implement but it was possible to achieve, reasonable results.  The baseline showed that the conditionally conjugate mean-field closed-form approach to VB, even though it has many simplifying assumptions, performed best both computationally and in terms of posterior accuracy \footnote{see Appendix 2 in the supplementary material}.  We feel a redesign of the Bayes LCA EHR phenotyping model using an exponential family variational Bayes mixture model with Stochatic Variational Inference (SVI) could potentially address these issues.  A closed-form approach also has the advantage that it can be published as an R/Python package, removing the need for clinical users to possess significant technical knowledge, perform time-consuming algorithm tuning needed for an automatic approach or create complex LCA models in \textit{Stan} modelling language.

\vspace{1cm}


\settocbibname{References}
\bibliographystyle{unsrt}
\bibliography{arxiv_paper}

\begin{thebibliography}{10}

\bibitem{boulanger703and}
Bruno Boulanger and Bradley~P Carlin.
\newblock How and why {B}ayesian statistics are revolutionizing pharmaceutical
  decision making.
\newblock {\em Clinical Researcher}, 703:20, 2021.

\bibitem{hripcsak2013next}
George Hripcsak and David~J Albers.
\newblock Next-generation phenotyping of {E}lectronic {H}ealth {R}ecords.
\newblock {\em Journal of the American Medical Informatics Association},
  20(1):117--121, 2013.

\bibitem{lanza2013latent}
Stephanie~T Lanza and Brittany~L Rhoades.
\newblock Latent class analysis: an alternative perspective on subgroup
  analysis in prevention and treatment.
\newblock {\em Prevention Science}, 14(2):157--168, 2013.

\bibitem{hubbard2019bayesian}
Rebecca~A Hubbard, Jing Huang, Joanna Harton, Arman Oganisian, Grace Choi,
  Levon Utidjian, Ihuoma Eneli, L~Charles Bailey, and Yong Chen.
\newblock A {B}ayesian latent class approach for {EHR}-based phenotyping.
\newblock {\em Statistics in {M}edicine}, 38(1):74--87, 2019.

\bibitem{dagenais2022use}
Simon Dagenais, Leo Russo, Ann Madsen, Jen Webster, and Lauren Becnel.
\newblock Use of real-world evidence to drive drug development strategy and
  inform clinical trial design.
\newblock {\em Clinical Pharmacology \& Therapeutics}, 111(1):77--89, 2022.

\bibitem{carpenter2017stan}
Bob Carpenter, Andrew Gelman, Matthew~D Hoffman, Daniel Lee, Ben Goodrich,
  Michael Betancourt, Marcus Brubaker, Jiqiang Guo, Peter Li, and Allen
  Riddell.
\newblock {S}tan: A probabilistic programming language.
\newblock {\em Journal of {S}tatistical {S}oftware}, 76(1), 2017.

\bibitem{hadjadj2004prognostic}
S~Hadjadj, D~Coisne, G~Mauco, S~Ragot, F~Duengler, P~Sosner, F~Torremocha,
  D~Herpin, and R~Marechaud.
\newblock Prognostic value of admission plasma glucose and {HbA1c} in acute
  myocardial infarction.
\newblock {\em Diabetic {M}edicine}, 21(4):305--310, 2004.

\bibitem{plummer2003jags}
Martyn Plummer et~al.
\newblock {JAGS}: A program for analysis of {B}ayesian graphical models using
  {G}ibbs sampling.
\newblock In {\em Proceedings of the 3rd international workshop on distributed
  statistical computing}, volume 124, pages 1--10. Vienna, Austria., 2003.

\bibitem{geman1984stochastic}
Stuart Geman and Donald Geman.
\newblock Stochastic relaxation, {G}ibbs distributions, and the {B}ayesian
  restoration of images.
\newblock {\em IEEE Transactions on pattern analysis and machine intelligence},
  (6):721--741, 1984.

\bibitem{yackulic2020need}
Charles~B Yackulic, Michael Dodrill, Maria Dzul, Jamie~S Sanderlin, and
  Janice~A Reid.
\newblock A need for speed in {B}ayesian population models: a practical guide
  to marginalizing and recovering discrete latent states.
\newblock {\em Ecological Applications}, 30(5):e02112, 2020.

\bibitem{dombowsky2020assessing}
Alexander Dombowsky.
\newblock {\em Assessing the quality of posterior samples from No-U-Turn
  Hamiltonian Monte Carlo}.
\newblock McGill University (Canada), 2020.

\bibitem{kucukelbir2017automatic}
Alp Kucukelbir, Dustin Tran, Rajesh Ranganath, Andrew Gelman, and David~M Blei.
\newblock Automatic differentiation variational inference.
\newblock {\em Journal of {M}achine {L}earning {R}esearch}, 2017.

\bibitem{kucukelbir2015automatic}
Alp Kucukelbir, Rajesh Ranganath, Andrew Gelman, and David Blei.
\newblock Automatic variational inference in {S}tan.
\newblock {\em Advances in {N}eural {I}nformation {P}rocessing {S}ystems}, 28,
  2015.

\bibitem{mcparland2016model}
Damien McParland and Isobel~Claire Gormley.
\newblock Model based clustering for mixed data: clust{MD}.
\newblock {\em Advances in Data Analysis and Classification}, 10(2):155--169,
  2016.

\bibitem{vehtari2015pareto}
Aki Vehtari, Daniel Simpson, Andrew Gelman, Yuling Yao, and Jonah Gabry.
\newblock Pareto smoothed importance sampling.
\newblock {\em arXiv preprint arXiv:1507.02646}, 2015.

\bibitem{magnusson2020leave}
M{\aa}ns Magnusson, Aki Vehtari, Johan Jonasson, and Michael Andersen.
\newblock Leave-one-out cross-validation for {B}ayesian model comparison in
  large data.
\newblock In {\em International Conference on Artificial Intelligence and
  Statistics}, pages 341--351. PMLR, 2020.

\bibitem{vehtari2017practical}
Aki Vehtari, Andrew Gelman, and Jonah Gabry.
\newblock Practical {B}ayesian model evaluation using leave-one-out
  cross-validation and {WAIC}.
\newblock {\em Statistics and {C}omputing}, 27(5):1413--1432, 2017.

\end{thebibliography}

\end{document}